\documentclass[letterpaper, 10 pt, conference]{ieeeconf}  % Comment this line out if you need a4paper

\IEEEoverridecommandlockouts                              % This command is only needed if 
\usepackage{graphicx}
\graphicspath{ {./Images/} }
\usepackage{graphics} % for pdf, bitmapped graphics files
\usepackage{epsfig} % for postscript graphics files
\usepackage{mathptmx} % assumes new font selection scheme installed
\usepackage{times} % assumes new font selection scheme installed
\usepackage{amsmath} % assumes amsmath package installed
\usepackage{amssymb}  % assumes amsmath package installed
\usepackage{float}
\usepackage{array,booktabs,arydshln,xcolor}
\newcommand\VRule[1][\arrayrulewidth]{\vrule width #1}
\usepackage{geometry}

\title{\LARGE \bf
Bode-based speed Proportional Integral and notch filter tuning of a Permanent Magnet Synchronous Machine driven flexible system
}

\author{Santiago~Ramos~Garces$^{1}$, Abdelmajid~Ben~Yahya, Nick~van~Oosterwyck, Dries~Jacques, Stijn~Derammelaere% <-this % stops a space
\thanks{$^{1}$ Santiago Ramos Garces is with the Department
of Electromechanics, University of Antwerp, Antwerp,
2020 Belgium
        {\tt\small Santiago.RamosGarces@student.uantwerpen.be}}%
}

\begin{document}
\newgeometry{top=50pt,left=46pt,right=46pt,bottom=74pt}

\maketitle
\thispagestyle{empty}
\pagestyle{empty}

%%%%%%%%%%%%%%%%%%%%%%%%%%%%%%%%%%%%%%%%%%%%%%%%%%%%%%%%%%%%%%%%%%%%%%%%%%%%%%%%
\begin{abstract}

A resonance and an antiresonance peak characterize many industrial mechanisms dynamics driven by a Permanent Magnet Synchronous Motor (PMSM). The presence of the resonance peak can lead to vibrations and instability of the system. On that account, advanced methods exist to tune the speed Proportional Integral (PI) controller based on adaptive or fuzzy theory. However, those methods require expertise in control theory and are not available in commercial drives. For that, this paper proposes a Bode-based method for PI parameters selection in combination with a notch filter that can be easily set in any industrial drive. The proposed method is compared with conventional tuning methods in a physical setup.\\

\end{abstract}

%%%%%%%%%%%%%%%%%%%%%%%%%%%%%%%%%%%%%%%%%%%%%%%%%%%%%%%%%%%%%%%%%%%%%%%%%%%%%%%%
\section{INTRODUCTION}
Permanent Magnet Synchronous Motors (PMSM) are extensively used in industrial motion control applications due to their high efficiency, accuracy, and fast dynamics. However, flexible coupling, shafts, and gearboxes are used to connect the load to the motor. Such couplings give rise to antiresonance and resonance frequencies. These can lead to torsional vibrations that can be transmitted to the load and cause damage to the structure or undesirable velocity and position errors that limit the tracking task's accuracy.\\
On that account, tracking a velocity or position reference becomes challenging because the controller action must be reduced to avoid vibrations, leading to lower bandwidth and poor performance. In literature, the most relevant control strategies are adaptive fractional order Proportional Integral Derivative (PID) \cite{Yu2016} and Proportional Integral (PI) \cite{Xie2018}, PI with H -infinity optimization of a modal model of the system \cite{bendrat2019speed}, fuzzy logic-based PI controller \cite{orlowska2004optimization},\cite{mokrani2003fuzzy}, two degrees of freedom PI controller based on generalized predictive control \cite{lu2013gpc}, optimal PID control \cite{Sencer2017} and PI controller combined with a disturbance observer and jerk feedforward \cite{chen2017composite}. Even though the addressed methods are robust and efficient, their implementation is complex, especially in commercially common industrial drives. Moreover, the techniques \cite{Xie2018}, \cite{bendrat2019speed}, \cite{lu2013gpc}, \cite{orlowska2004optimization}, \cite{chen2017composite} still require the application of specific tuning by the control engineer.\\% Nevertheless, some methods require a system model, external sensors, or high computational effort to be implemented.
On the contrary, if the system's inertia is fixed, a more straightforward control algorithm that utilizes a notch filter in the loop to compensate the resonance frequency in combination with a PI controller is valuable. Authors in \cite{Yang2014}, \cite{Bahn2017}, \cite{Schmidt1999} proposed an adaptive notch filter through different methods for estimation of the resonance frequency but did not focus on tuning the PI parameters. Besides, adaptive algorithms require extra computational effort and are not available in the conventional cascade loop of industrial drivers for PMSM. In this paper, a deterministic method to tune a PI controller and notch filter combination that can be used in any industrial drive is proposed to improve the performance compared to conventional tuning methods. It determines the controller gains based on a nonparametric identification of the system in the frequency domain that only required measurement data from the PMSM encoder. The paper is structured as follows; In Section~\ref{sec:1}, the modeling and identification of the system are explained. Section~\ref{sec:2} presents the control law formulation and simulation. Finally, in Section~\ref{sec:3}, implementation results on an actual setup with two different structural flexibilities will be presented to compare the performance of conventional methods with the proposed controller.

\section{Modeling of the system dynamics}\label{sec:1}
\begin{figure}[hbtp]
	\centering
	\includegraphics[scale=0.6]{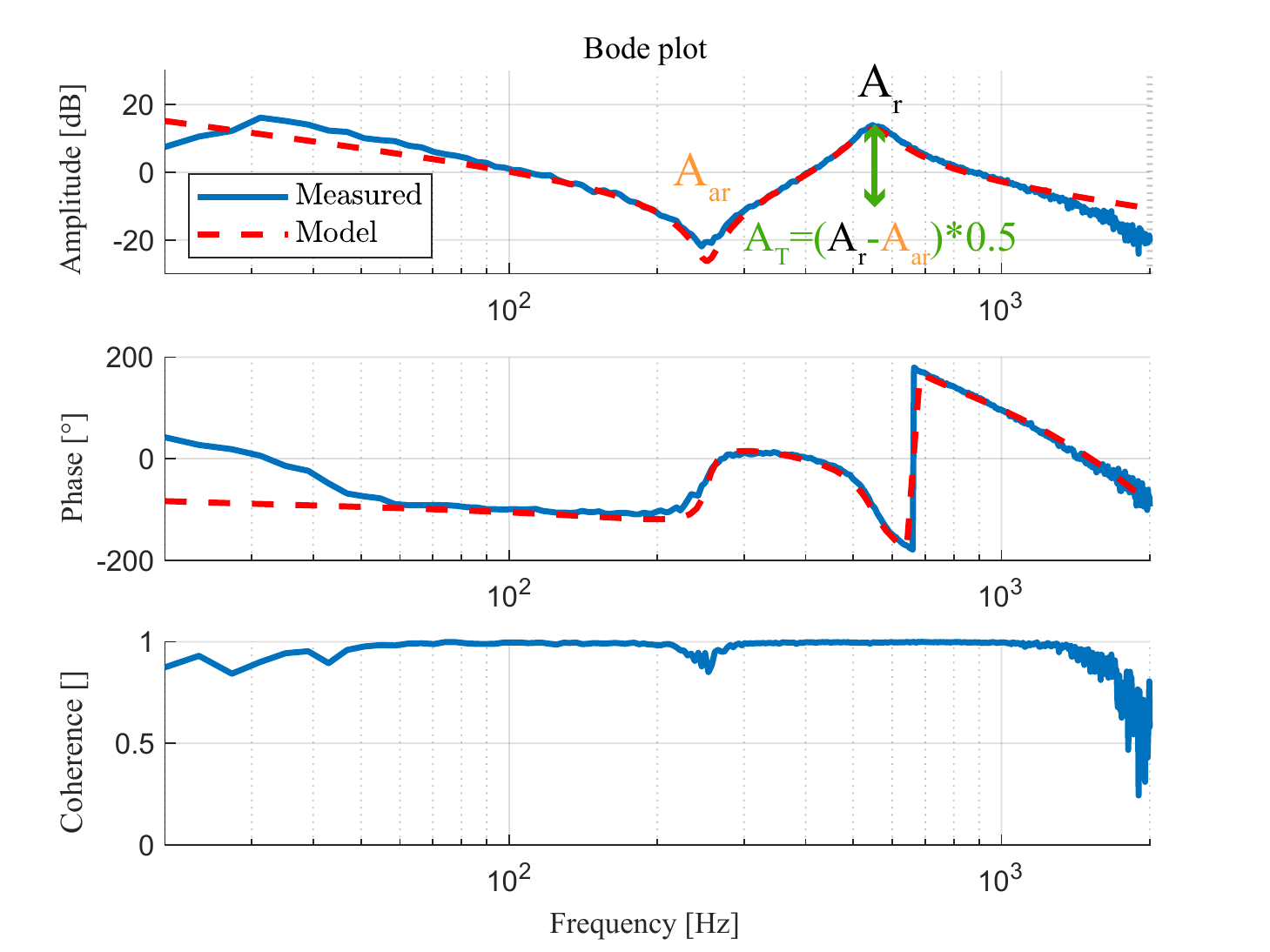}
	\caption{Bode of the modeled and the measured system }
	\label{fig:BodeLin}
\end{figure}

Most motion control applications consist of a mechanism driven by a motor through a mechanical coupling. The system exhibits some structural flexibility that depends on the stiffness and damping coefficients of the coupling. Furthermore, many of the mechanisms to be controlled in motion control applications consist of multiple components, and the mathematical model using physical laws can be hard to obtain. For that reason, the main idea with the proposed tuning method is to use a nonparametric model of the system that can be computed efficiently with experimental data. The input is white noise applied to the torque of the PMSM, and the output is the motor's shaft velocity through the attached encoder. This configuration is the so-called the collocated case \cite{10.5555/550726} because the sensor and actuator are located in the same place. Then, the Bode diagram is estimated using Welch's averaged modified periodogram method.\\
Even though the proposed controller does not need a mathematical model to be obtained, in Section~\ref{sec:2.2}, simulations are used to justify the selection of the notch filter's depth. On that account, a rotatory two-mass spring-damper system \cite{doi:10.1177/14644193211019943}, \cite{vanbecelaere2020online}, \cite{li2021linear} is used to represent the model, and for simplification, the friction is considered linear \cite{vanbecelaere2020online}. In order to validate the theoretical model, its Bode diagram is compared with the Bode diagram of the setup used for implementation in Section~\ref{sec:3}, as depicted in Fig.~\ref{fig:BodeLin}. The coherence in Fig.~\ref{fig:BodeLin} is a frequency dependant function with values between 0 and 1 that indicate how correlated the input and the output are at each frequency \cite{pintelon2012system}.

\section{Controller Formulation}\label{sec:2}
\begin{figure}[hbtp]
	\centering
	\includegraphics[width=1.0\linewidth]{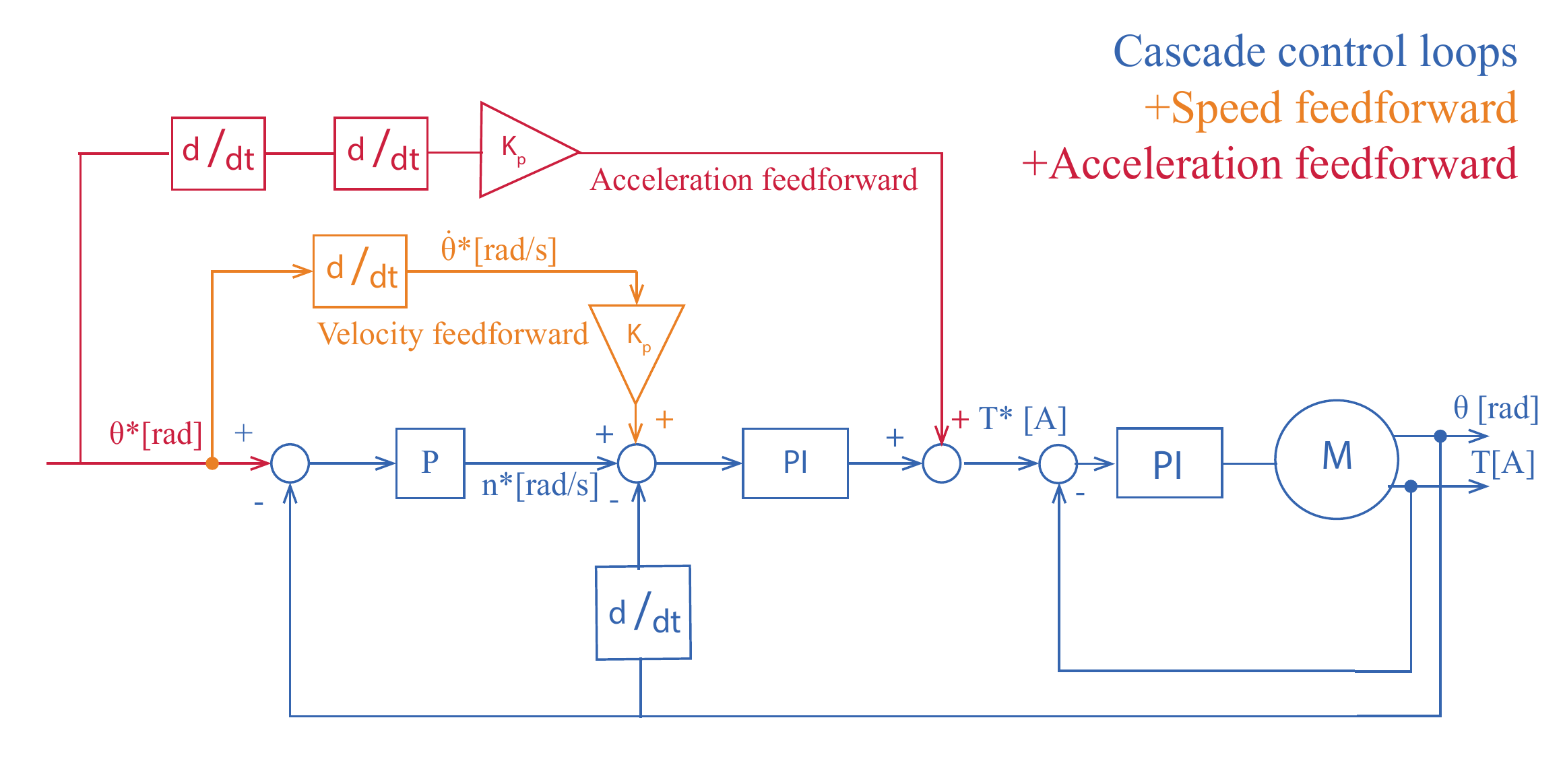}
	\caption{Cascade control loop for motion control applications}
	\label{fig:ConLoop}
\end{figure}

The conventional control loop of motion control applications implemented in the lion's share of the industrial driver's software like Siemens S120 and Beckhoff AX series is depicted in Fig.~\ref{fig:ConLoop}. The current controller is usually tuned by the supplier of the motor, based on exact knowledge of the motor coil characteristics. However, the optimal velocity and position control settings strongly depend on the dynamics of the load. The system's dynamic characteristics to control consist of an antiresonance and a resonance peak (Fig.~\ref{fig:BodeLin}). The presence of that resonance peak limits the controller's bandwidth. An increase in the proportional gain can lead to an unstable condition based on the simplified Nyquist criterion \cite{richard2008modern}, and the system's performance is poor. The rest of this section introduces the proposed tuning procedure step by step.\\
\subsection{Tuning Procedure}\label{sec:2.2}
\begin{figure}[H]
	\centering
	\includegraphics[width=0.7\linewidth]{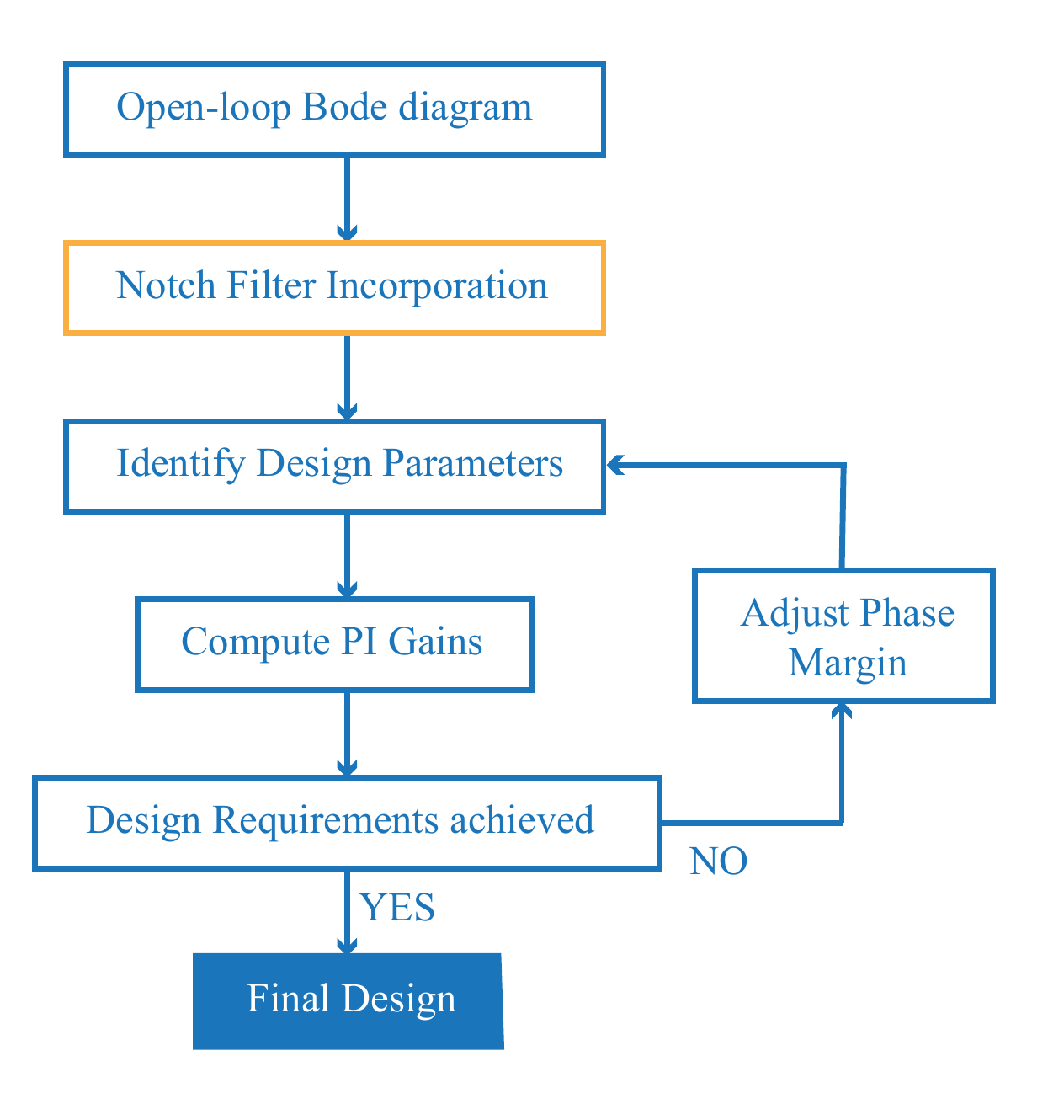}
	\caption{Flow diagram of the controller method design}
	\label{fig:DContr}
\end{figure}

\begin{figure}[hbtp]
	\centering
	\includegraphics[width=0.6\linewidth]{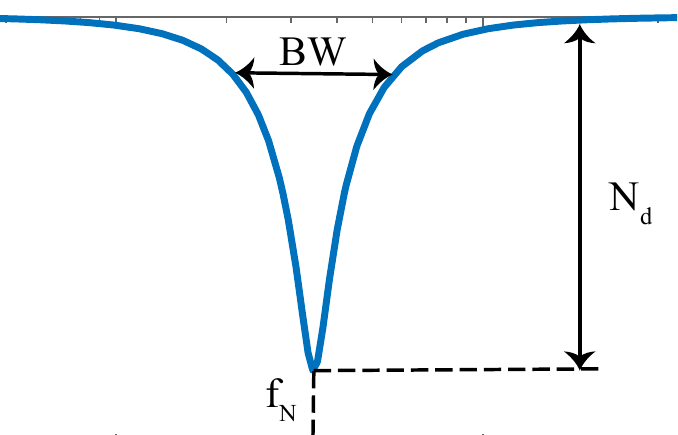}
	\caption{Graphical representation of the Notch Filter}
	\label{fig:Notch}
\end{figure}

\begin{figure}[H]
	\centering
	\includegraphics[width=0.9\linewidth]{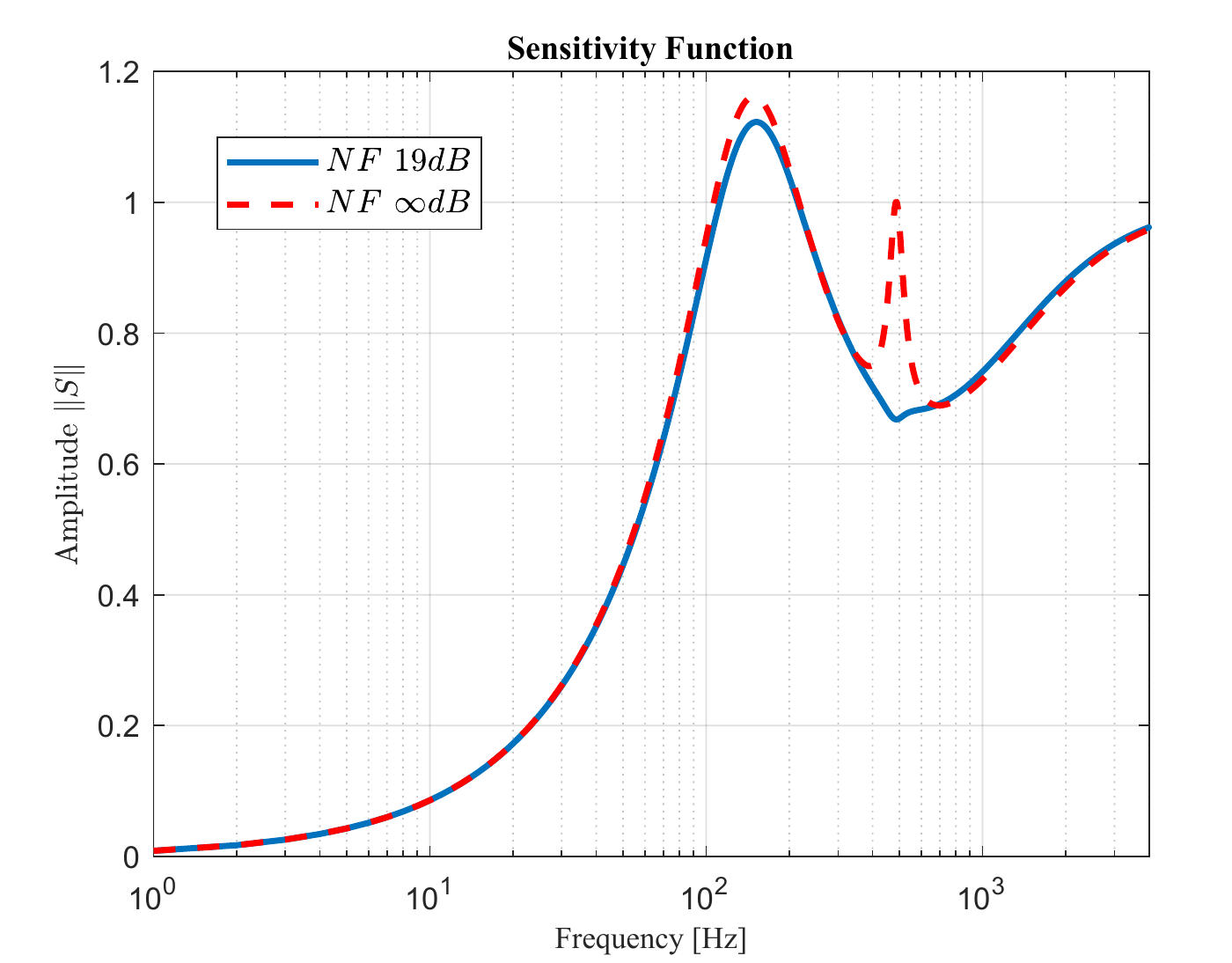}
	\caption{Sensitivity function of the theoretical system with an infinite and finite notch depth}
	\label{fig:Sen}
\end{figure}

\begin{figure}[H]
	\centering
	\includegraphics[width=1.0\linewidth]{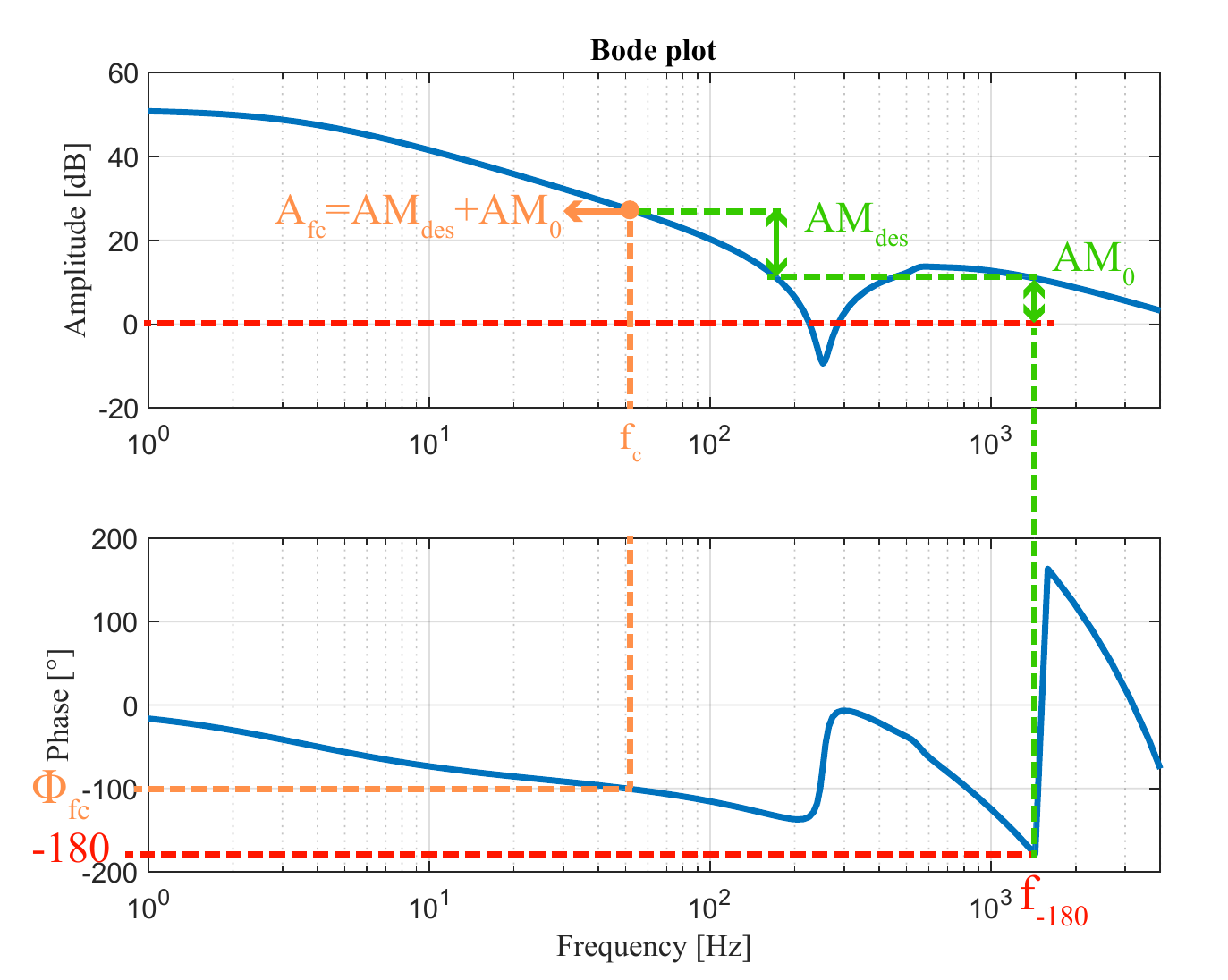}
	\caption{Bode plot including the Notch filter for controller design}
	\label{fig:CD}
\end{figure}

The proposed procedure consists of six steps, as shown in Fig.~\ref{fig:DContr}. The initial step requires the system's Bode diagram, which is the only necessary input for this method.
The open-loop Bode diagram that relates the motor's torque or current to the velocity of the PMSM shaft as an output can be obtained as mentioned in section~\ref{sec:1}. \\
The second step of the method is the implementation of a notch filter. This tool is crucial for the proper setting of the controller. Indeed, the main functionality of the filter is to suppress the effect of the resonance frequency of the mechanical system. 
Compensating for the resonance peak allows an aggressive control action and better performance, as demonstrated in section~\ref{sec:3} with the real system. The notch filter is a narrow-band filter that can be implemented as a second-order difference equation.\\ 
However, setting up the filter can be done by selecting three parameters, as depicted in Fig.~\ref{fig:Notch}. Those parameters (Fig.~\ref{fig:Notch}) are the notch frequency $f_{N}$, bandwidth $BW$, and notch depth $N_{d}$. The notch frequency can be identified from the Bode diagram as it must have the same frequency as the resonance peak. On the other hand, the bandwidth is not that straightforward to select, but based on the authors' recommendation in \cite{AG2012}, choosing between $BW=f_{N}$ to $BW=2f_{N}$ of the notch frequency is ideal.
\\ Finally, the depth of the filter plays an important role, and its selection is analyzed here in detail because there are two main options. Either an infinite depth or half the difference between resonance and antiresonance peak ($A_{T}$) from Fig.~\ref{fig:BodeLin} as suggested in \cite{AG2012}. 
The first option simplified setting up the filter reducing the number of parameters to two. However, the infinite depth has implications on the system's stability. 
Fig.~\ref{fig:Sen} depicts the cascade notch filter-process sensitivity function ($S(j\omega)$) for the theoretical rotatory two-mass spring-damper system with infinite and finite depth. \\
The sensitivity function reveals that the system with infinite depth is more sensitive to uncertainties for the frequencies in the vicinity of the notch and can lead to unstable behavior. Meanwhile, the finite depth allows the system to have better robustness to changes in the system that are not modeled and keep a better stability condition proven by the lower sensitivity compared with the infinite depth counterpart. Thus, the advantage of the finite depth's notch filter is considered a better solution in this paper.\\

\begin{subequations}
	\begin{align}
		C(s)=K_{p}(\frac{T_{i}s+1}{T_{i}s}) \label{eqn:line-1}\\
		C(j\omega)=K_{p}(\frac{T_{i}j\omega+1}{T_{i}j\omega}) \label{eqn:line-2}\\
		\left|C(j\omega)\right|=20\log_{10}\left|K_{p}\right|+20\log_{10}\left|\frac{j\omega T_{i}+1}{j\omega T_{i}}\right| \label{eqn:line-3}\\
		\angle C(j\omega)=\tan(\omega T_{i}) - 90 
	\label{eqn:all-lines}
	\end{align}
\end{subequations}
The next step in the method's flow diagram is identifying the necessary parameters to compute the PI gains denoted as $K_{p}$ and $T_{i}$. The PI controller is formulated as (\ref{eqn:line-1}), which becomes (\ref{eqn:line-2}) in the frequency domain. Equation (\ref{eqn:line-2}) can be denoted as the amplitude (\ref{eqn:line-3}) and phase (\ref{eqn:all-lines}). After including the Notch filter, the PI settings are based on the Bode diagram from notch input to speed output, and its selection is based on the desired amplitude margin $(AM)$ and phase margin $(PM)$. The amplitude and phase margin play an essential role in the system's stability and robustness. Furthermore, the phase margin is associated with the system's transient behavior. Indeed, the higher the phase margin, the better the transient response because of the relation $\zeta = PM/100^{\circ}$ \cite{10.5555/516039}, where $\zeta$ refers to the damping coefficient. Even though the previous assumption is only valid for second-order systems, higher-order dynamics can be approximated as a second-order system. In this case, the amplitude margin must be at least $10 dB$ as advised in \cite{AG2012} so that the system is stable and the control action can still be powerful enough.\\
Regarding the phase margin, a value between $[30^{\circ} \cdots  60^{\circ}]$ is suitable for ensuring an acceptable transient response \cite{Astrom:1514097}. However, if the required $\zeta$ is specified, the authors in \cite{nise2020control} proposed to compute the desired phase margin as in (\ref{Eq20}).\\ 
\begin{equation}\label{Eq20}
     \phi_{PM{des}} =\tan^{-1}\left (\frac{2 \zeta}{\sqrt{\sqrt{1+4\zeta^{4}}-2 \zeta^{2}}}\right )
\end{equation}
The initial amplitude margin $AM_{0}$ must be read on the magnitude plot at the frequency where the phase crosses $\mbox{--}180^{\circ}$ ($f_{ \mbox{--}180}$ in Fig.~\ref{fig:CD}). Then, the frequency denoted as $f_{c}$ for which the system has an amplitude equal to the initial amplitude margin plus the desired one ($AM_{0}+AM_{des}$) must be identified on the magnitude plot. Finally, the phase $\phi_{f_{c}}$  at the frequency $f_{c}$ must be read from the phase plot. The open-loop (a cascade of the controller - notch filter - process) frequency response can be shaped by adding a PI controller. The PI controller is designed so that the magnitude of the open-loop is $0~dB$ at this $f_{c}$ based on the principle that the Nyquist curve can be moved to an arbitrary desired point \cite{Astrom:1514097}.\\
 In that case, the desired amplitude margin $AM_{des}$ will be obtained. This is the case if the following condition is met:\\
\begin{equation}\label{Eq9}
		20\log_{10}\left|K_{p}\right|+20\log_{10}\left|\frac{j \cdot 2\pi f_{c}T_{i}+1}{j \cdot 2\pi f_{c}T_{i}}\right|+A_{f_{c}}= 0 
\end{equation}
under the assumptions from  (\ref{Eq13}).
\begin{equation}\label{Eq13}
\begin{array}{r@{}l}
    \angle \left(\frac{j \cdot 2\pi f T_{i}+1}{j \cdot 2\pi f T_{i}}\right)_{f=f_{-180}} = 0 \\ \\
    \left|\frac{j \cdot 2\pi f T_{i}+1}{j \cdot 2\pi f T_{i}}\right|_{f=f_{-180}}=0
\end{array}
\end{equation}
Moreover, to obtain the required phase margin, an additional requirement is added:
\begin{equation}\label{Eq10}
    \phi_{f_{c}}-90+\tan(2\pi f_{c}T_{i})=-180+\phi_{PM{des}}
\end{equation}
Equation (\ref{Eq10}) leads to the gain from  (\ref{Eq11}), and then replacing this value in  (\ref{Eq9}), the $K_{p}$ from  (\ref{Eq12}) is obtained.%where $K_{p}$ and $T_{i}$ refer to the proportional and integral gains of the $PI$ controller. Then, the main idea is to establish $f_{c}$ as the new cut-off frequency using the magnitude of the $PI$ controller and the phase for achieving the desired phase margin $PM_{des}$. For that, the magnitude at the design frequency $f_{c}$ must be zero, and the phase equals $-180+\phi_{PM{des}}$. The previous condition is represented by the system of equations from  (\ref{Eq9}), whose solution leads to the gains from  (\ref{Eq10}).\\
\begin{equation}\label{Eq11}
		T_{i}=\frac{\tan(-180+\phi_{PM{des}}-\phi_{f_{c}}+90)}{2\pi f_{c}}
\end{equation}
\begin{equation}\label{Eq12}
		K_{p}=10^{-\frac{A_{f_{c}}+20\log_{10}\left|\frac{j \cdot 2\pi f_{c}T_{i}+1}{j \cdot 2\pi f_{c}T_{i}}\right|}{20}} 
\end{equation}
The last step of the proposed method is to verify that the system satisfies the design criteria with the addition in the loop of the PI controller with the assessed gains. The Bode diagram of the open-loop, including the plant, notch filter, and controller, is required for that purpose. Suppose the system does not fulfill the requirements. In that case, the procedure must be repeated from the third step but adding an offset to the desired phase margin $(\sim3^{\circ})$ to compensate for the added error by the controller's phase graph. The above mention occurs because %in  (\ref{Eq9}), $AM_{0}$ was obtained from the Bode in Fig.~\ref{fig:CD}, assuming that the contribution of the controller in phase at the frequency at which the system reaches $-180^{\circ}$ denoted as ($f_{-180}$) is zero. However,
the phase of the PI controller goes asymptotically to zero for high frequencies but is not precisely zero at $f_{\mbox{--}180}$, adding an error to the system \cite{10.5555/516039}.\\ 

\section{Implementation and Results}\label{sec:3}
\begin{figure}[H]
	\centering
	\includegraphics[width=1.0\linewidth]{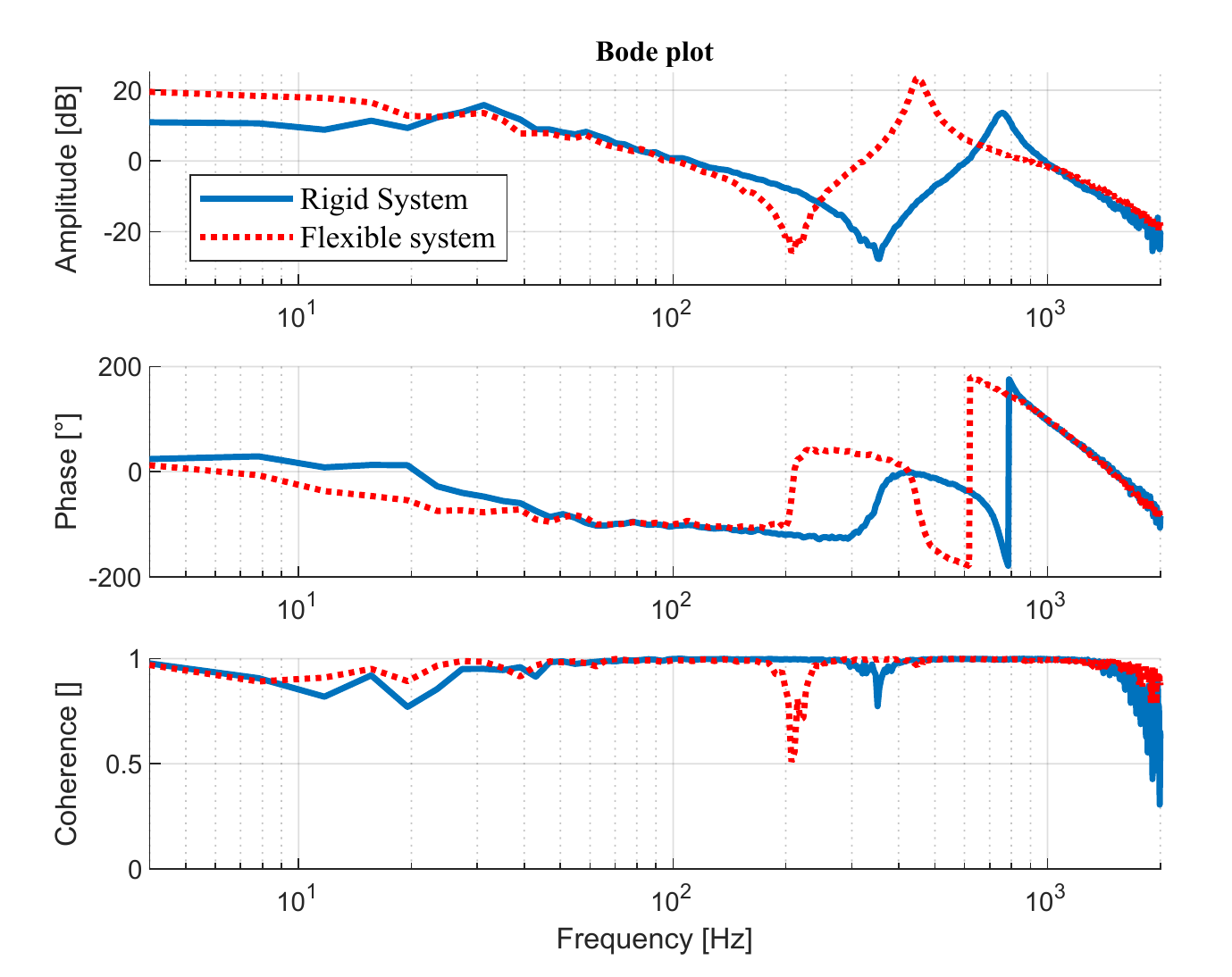}
	\caption{Open-loop Bode diagram for the rigid and flexible system}
	\label{fig:OLR}
\end{figure}

\begin{figure}[H]
	\centering
	\includegraphics[width=0.97\linewidth]{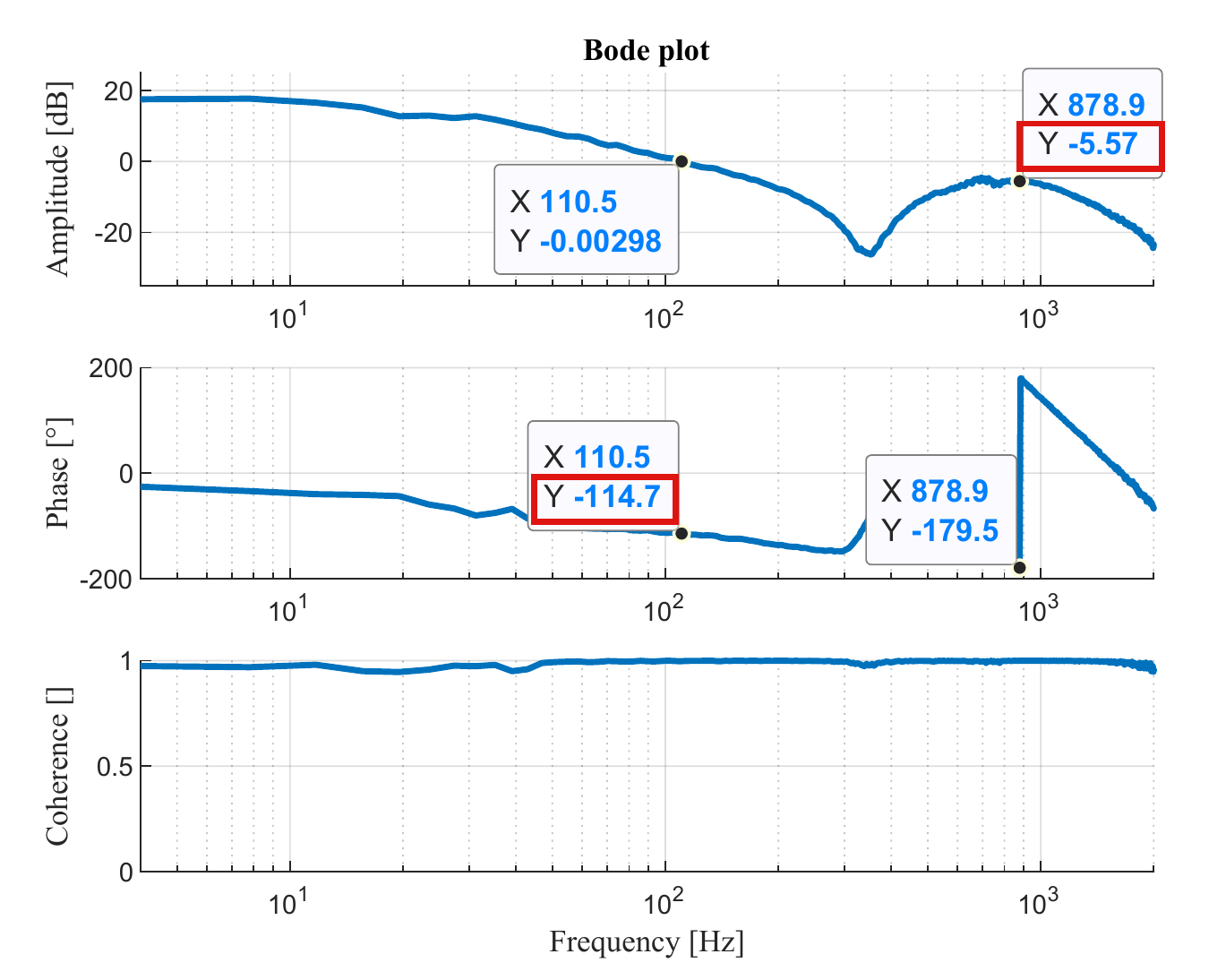}
	\caption{Bode diagram for controller validation for the rigid system}
	\label{fig:BRV}
\end{figure}

\begin{figure}[H]
	\centering
	\includegraphics[width=0.98\linewidth]{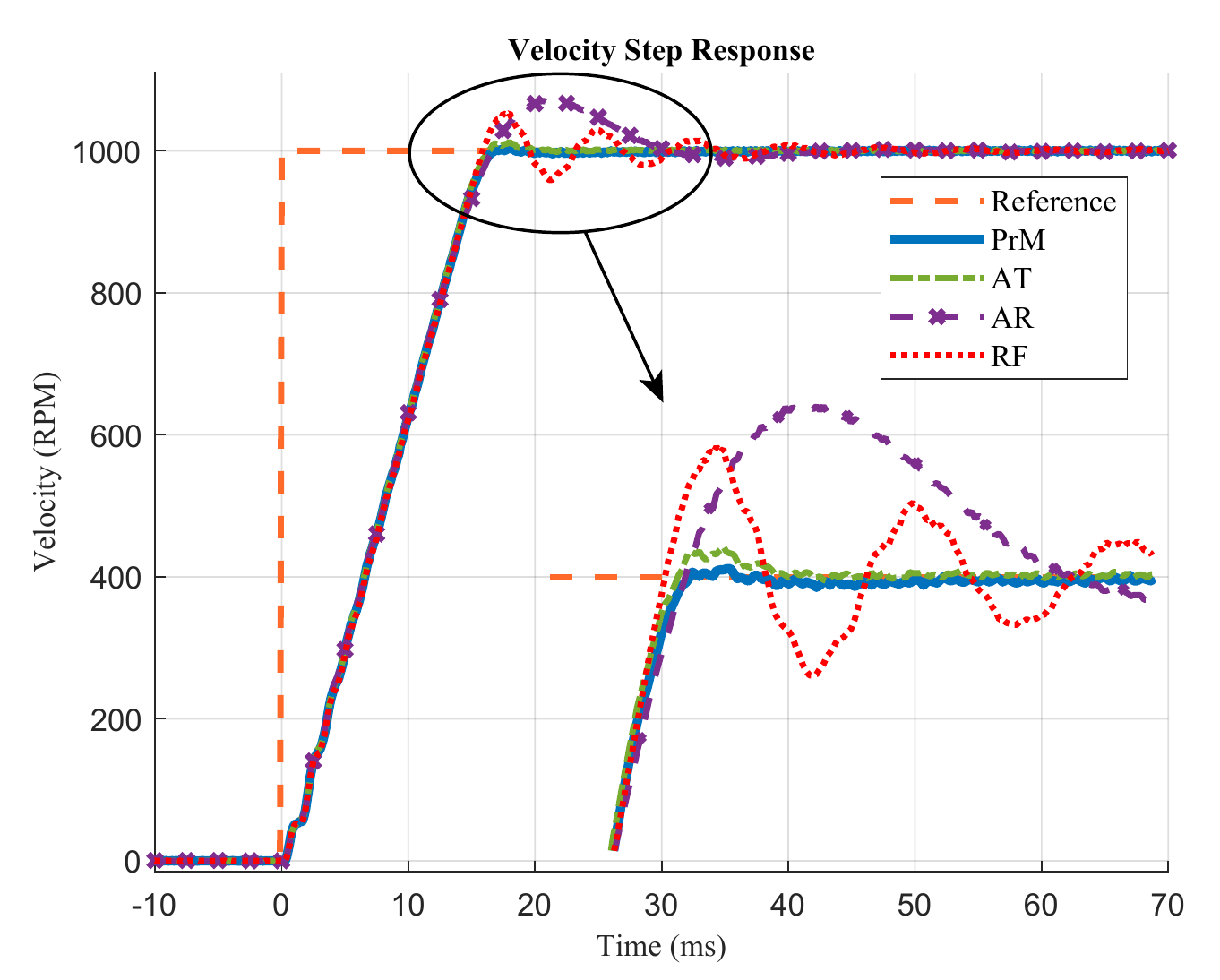}
	\caption{Velocity step response of the rigid system with the four different controllers }
	\label{fig:VSR}
\end{figure}

To validate the approach of the previous section, a Siemens set up with a PMSM $1FK7042-2AF71-1QG0$ as controlled motor, an Asynchronous motor $1LE1003-0DA22-2AB4$ as load, a  control unit $6SL3040-1MA010AA0 $ as the PMSM controller, and a Rotex coupling $GS-19 AL-H/GS-24 AL-H$ is used.\\
\begin{figure}[H]
	\centering
	\includegraphics[width=0.5\linewidth]{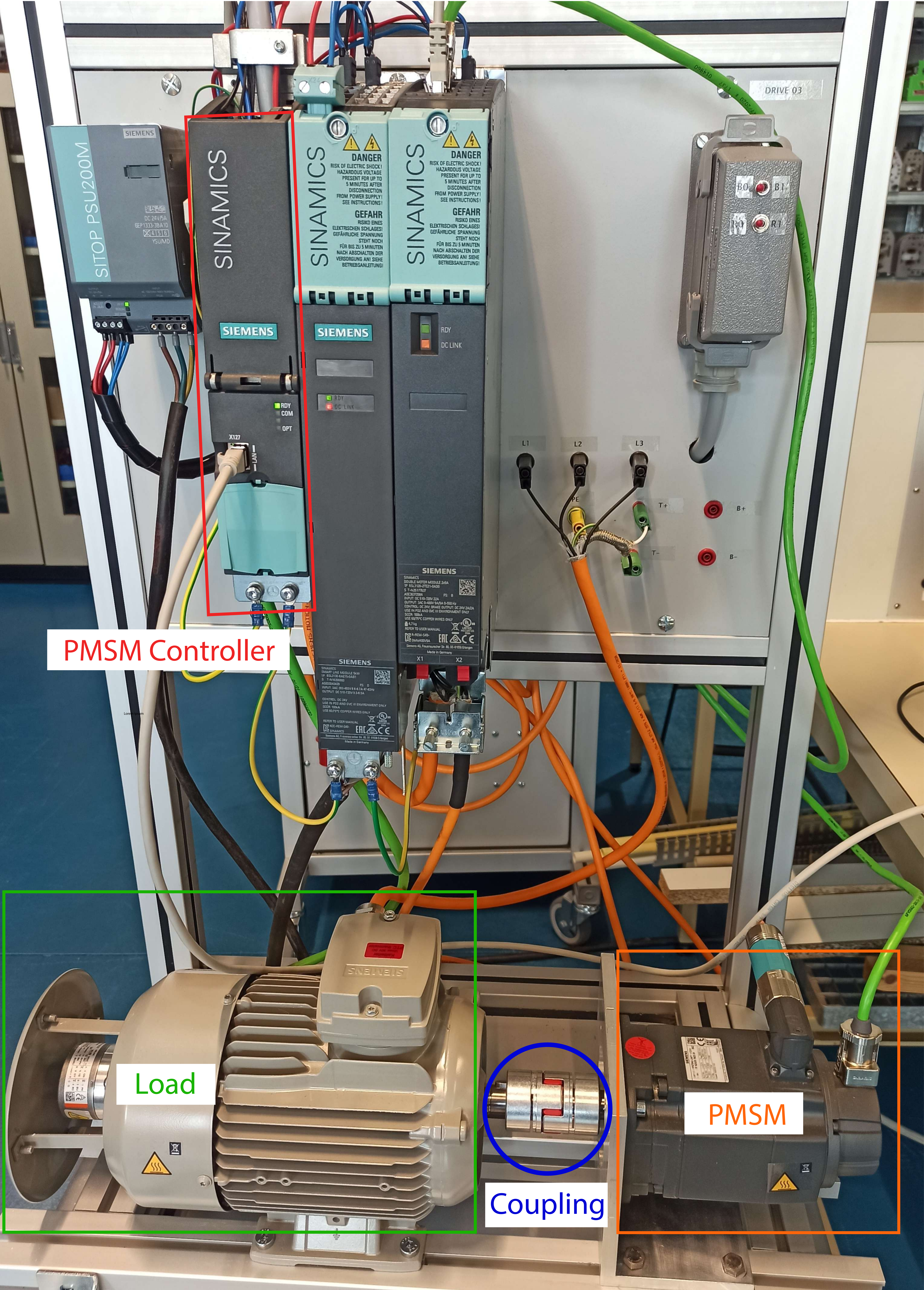}
	\caption{Hardware for implementation}
	\label{fig:HI}
\end{figure}
The coupling (Fig.~\ref{fig:HI}) consists of two aluminum hubs held together through a flexible spider made of polyurethane. In this section, the performance of the speed PI controller tuned as in Section~\ref{sec:2} is compared with controllers design according to other conventional frequency-based methods like Ziegler and Nichols based on relay feedback \cite{Astrom:1514097}, antiresonance frequency \cite{van2019cad}, and the autotune function available in the controller software \cite{AG2012}. Two different types of spiders will be tested to modify the stiffness and damping of the coupling. The spiders will be referred to as rigid and flexible. The system dynamics for the rigid and flexible spider can be observed in Fig.~\ref{fig:OLR}. \\
From this Bode diagram and the application of the mathematical formulas described in \cite{AG2012}, the calculated stiffness and damping coefficients for the rigid system are $k = 5118~Nms/rad$, $b = 0.117~Nm/rad$, and $k = 1828~Nms/rad$, $b = 0.049~Nm/rad$ for the flexible system.\\
For the rigid system controller design, the autotune method results in an amplitude margin of ($5.4~dB$), also used as a design criterion for the proposed method. Regarding phase margin, a value of $65^{\circ}$ is selected. For the experiment, only the finite notch depth variant is considered due to its advantage in the sense of stability. The controller gains, notch filter, and performance parameters obtained by the tuning methods are listed in Table~\ref{tab4}. The notation (PrM) refers to the proposed method, (AR) to antiresonance, (RF) to relay feedback, and (AT) to autotune. Furthermore, the Bode diagram including the PI controller for validation of the proposed controller is depicted in Fig.~\ref{fig:BRV}, and the step velocity response is shown in Fig.~\ref{fig:VSR}, respectively. \\
First, from Fig.~\ref{fig:BRV} and Table~\ref{tab4}, the proposed controller fulfills the design criteria with some errors for the phase margin ($0.3^{\circ}$) that can be neglected. The overshoot, settling time ($2\%$ criterion), and Integral of Time multiplied by the Absolute Error (ITAE) index \cite{graham1953synthesis} indicate the performance of each tuning method. The latter shows that the AR method produces the worst results, especially for the overshoot and ITAE with $7.1\%$ and $480.6$, respectively. In addition, the RF method presents an acceptable overshoot of $5.4\%$, but the settling time and ITAE index are significantly longer compared to AT.\\
 \begin{table}[H]
	\begin{center}
	   	\caption{\label{tab4}  Performance criteria and setting parameters for the rigid system}
	\begin{tabular}{!{\VRule[1pt]}c!{\VRule[1pt]}c!{\VRule[1pt]}c!{\VRule[1pt]}c!{\VRule[1pt]}c!{\VRule[1pt]}c}
\specialrule{1pt}{0pt}{0pt}
\textbf{}                   & \textbf{PrM} & \textbf{AR} & \textbf{RF} & \textbf{AT} \\ \specialrule{1pt}{0pt}{0pt}
\textbf{$K_{p}~[Nms/rad]$}       & 0.992        & 0.369       & 0.494       & 1.05       \\ 
\textbf{$T_{i}~[ms]$}            & 24.58       & 4.69        & 1.02        & 8.39       \\ \specialrule{1pt}{0pt}{0pt}
\textbf{$N_{f}~[Hz]$}     & 750         & 750         & -         & 753         \\ 
\textbf{$BW~[Hz]$}     & 750         & 750         & -         & 761         \\ 
\textbf{$N_{d}~[dB]$}   & 20          & 20          & -          & $\infty$    \\ \specialrule{1pt}{0pt}{0pt}
\textbf{$PM~ [^{\circ}]$}            & 65.3        & 55.5        & 11.62            & 57           \\ 
\textbf{$AM~[dB]$}            & 5.57       & 13.78       & 5.6           & 5.4          \\ 
\textbf{$BW_{C}~[Hz]$}            & 169       & 86.4       & 155           & 179          \\ \specialrule{1pt}{0pt}{0pt}
\textbf{$Overshoot~[\%]$}     & 0.3         & 7.1         & 5.4          & 1.2           \\ 
\textbf{$Settling~time~[ms]$} & 15.8        & 27.7          & 28.6           & 15.66           \\
\textbf{$ITAE~index$}            & 362.1       & 480.6       & 543.3           & 356.8          \\ \specialrule{1pt}{0pt}{0pt}

\end{tabular}
 
	\end{center}
\end{table}
 Finally, PrM and AT achieve a short settling time and similar ITAE index. However, the best performance can be attributed to PrM since it has almost no overshoot ($0.3\%$), as depicted in Fig.~\ref{fig:VSR}. As mentioned in section~\ref{sec:2}, the phase margin plays a vital role in the transient response, but also, the bandwidth of the closed-loop system $(BW_{C})$ is relevant for the settling time. PrM achieves the highest phase margin and has a high bandwidth ensuring a low overshot and fast response. However, in the case of RF, even though the bandwidth is the highest, the phase margin is too low, producing oscillatory behavior.\\

\begin{figure}[H]
	\centering
	\includegraphics[width=0.96\linewidth]{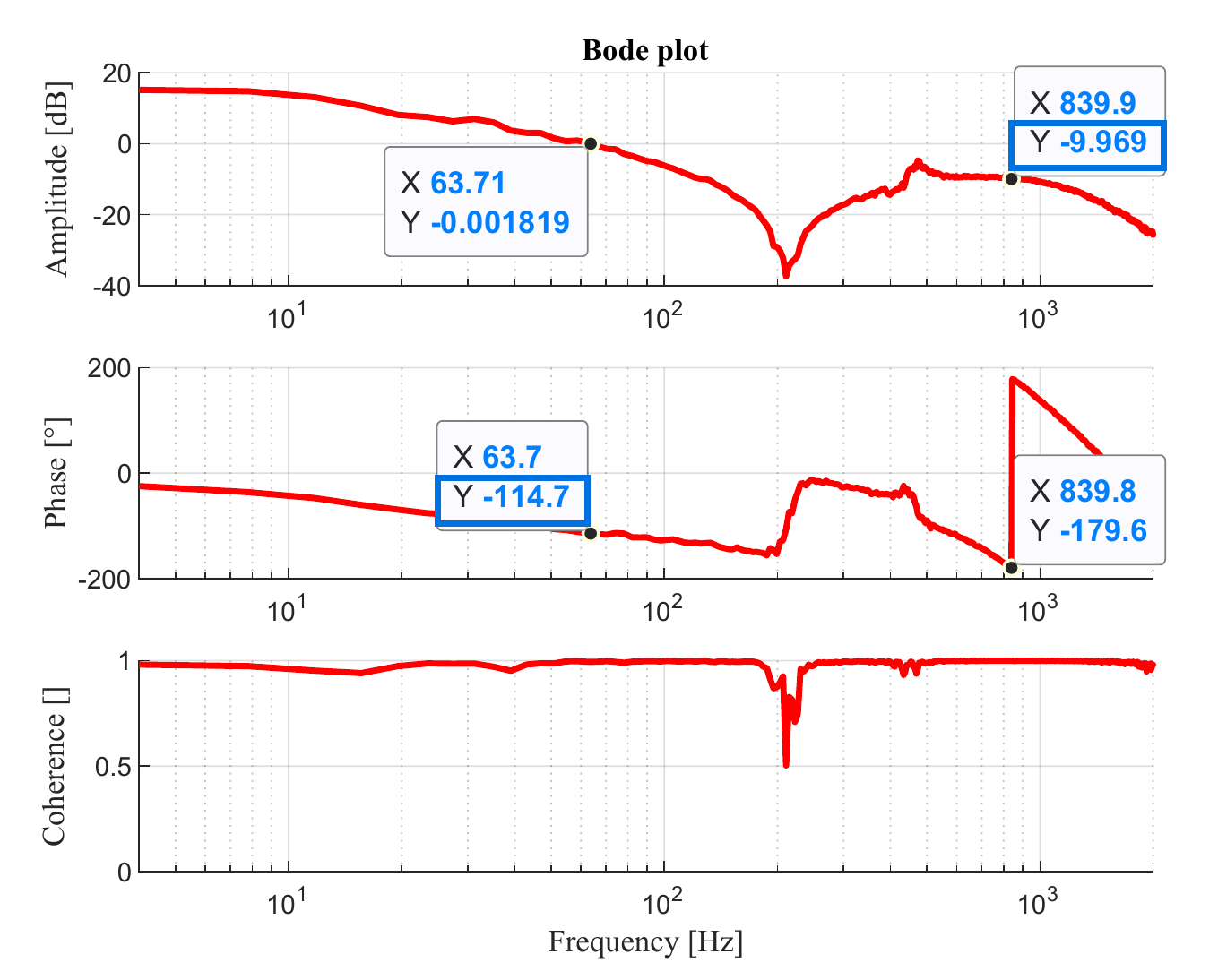}
	\caption{Bode diagram for controller validation for flexible system}
	\label{fig:BFV}
\end{figure}

\begin{figure}[H]
	\centering
	\includegraphics[width=0.96\linewidth]{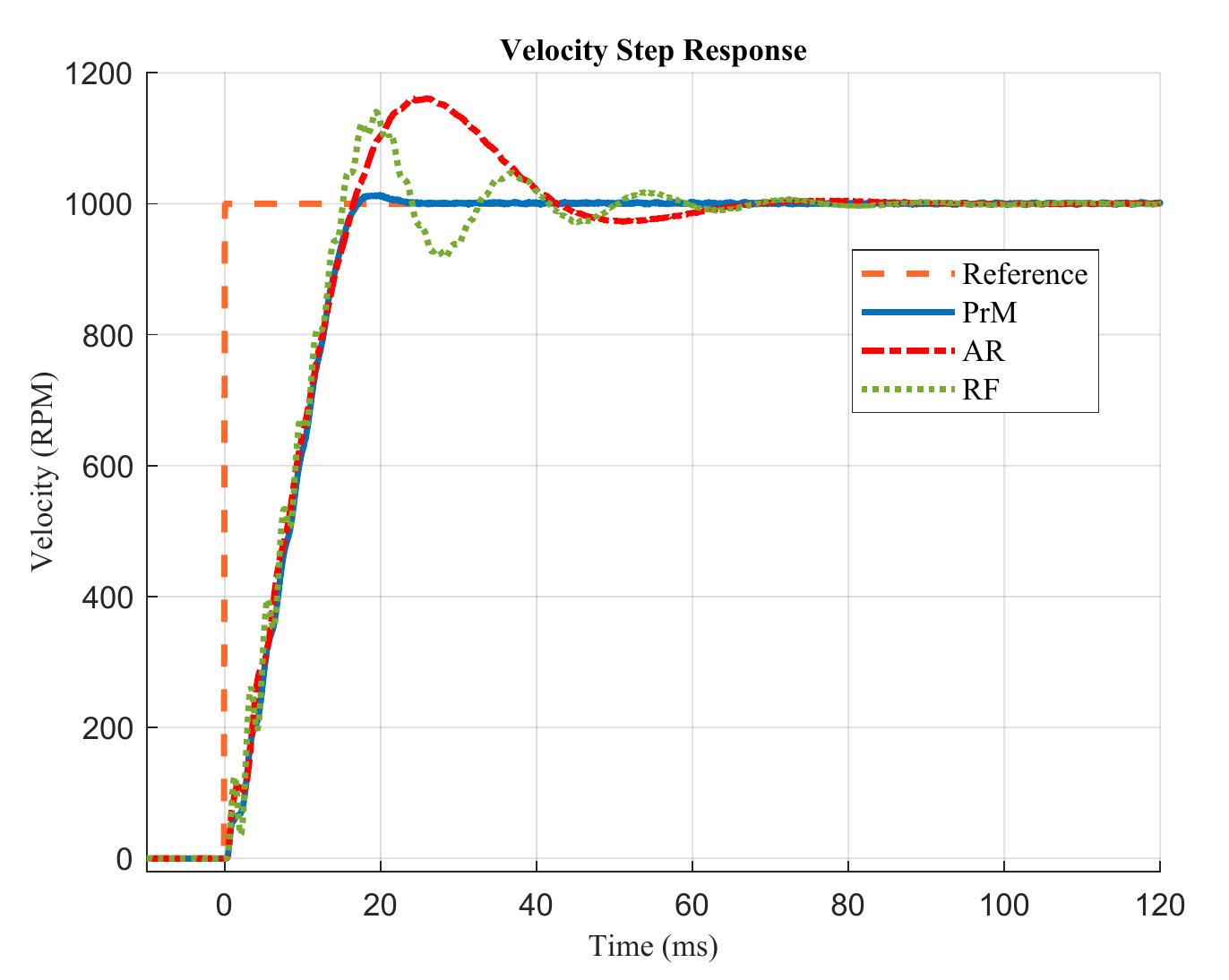}
	\caption{Velocity step response of the flexible system with the proposed controller}
	\label{fig:VSF}
\end{figure}

\begin{figure}[H]
	\centering
	\includegraphics[width=0.96\linewidth]{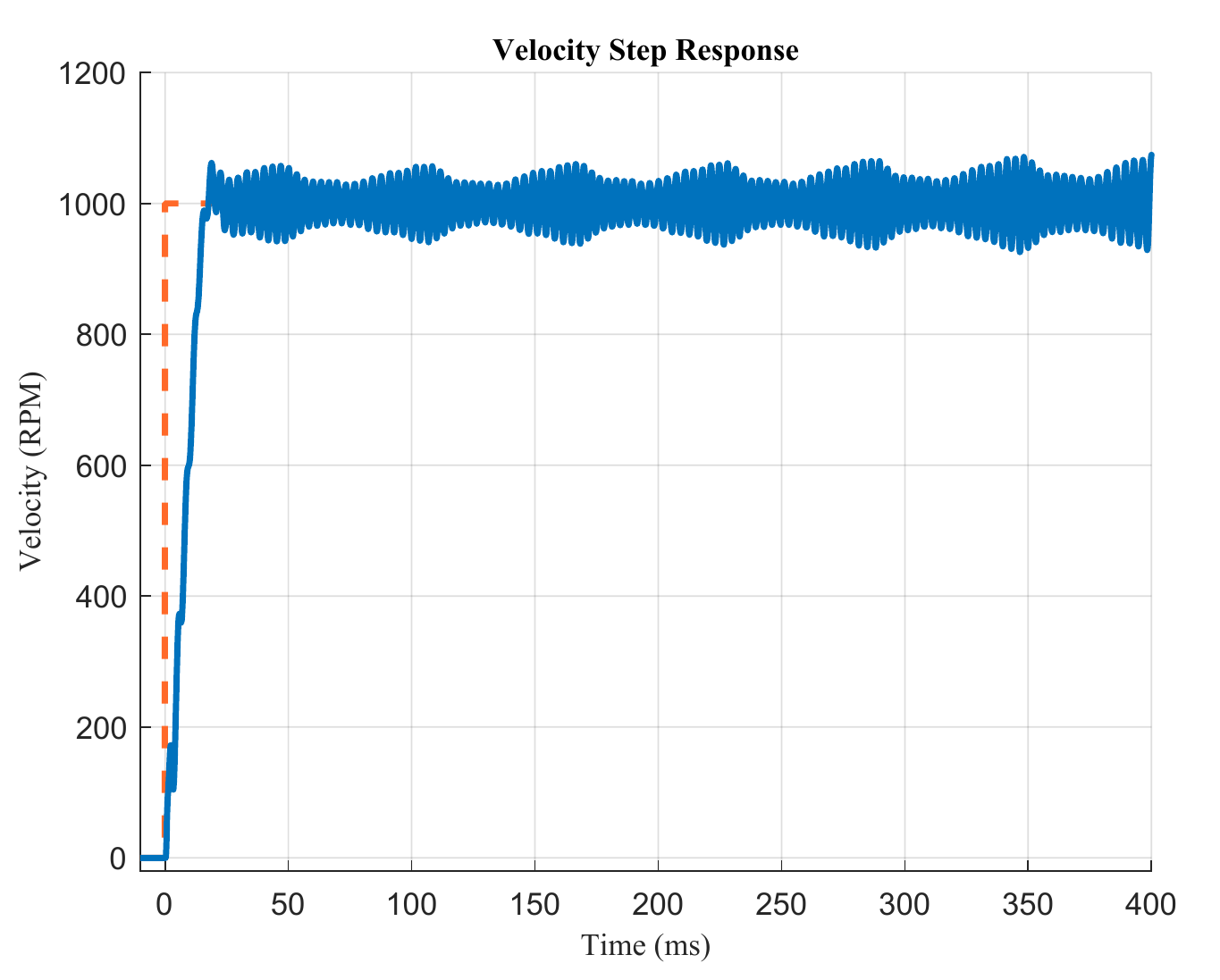}
	\caption{Velocity step response of the flexible system with the autotune controller}
	\label{fig:VSA}
\end{figure}

 On the other hand, AT generates a controller that makes the system unstable for the flexible spider. Thus, the comparison will be performed between PrM, AR, and RF. In Table~\ref{tab5}, the relevant results for all controllers are summarized. Finally, Fig.~\ref{fig:BFV} depicts the Bode diagram with the inclusion of the designed controller for verification of design parameters' fulfillment, and Fig.~\ref{fig:VSF} shows the velocity step response with the control alternatives. \\
   The designed controller fulfills the design criteria for the flexible system.%, with some acceptable errors again for the phase margin. The error is above the desired phase margin, which contributes to better performance. 
   Furthermore, it is observed that PrM has the highest bandwidth. Consequently, it achieves the shortest settling time ($16.1~ms$) and the lowest ITAE index ($391.8$) and overshoot ($1.35~\%$) compared to its counterparts, as observed in Fig.~\ref{fig:VSF}.\\
  Regarding AR and RF alternatives, similar results as with the rigid system are evidenced, showing a slight difference for the overshoot. Besides, AR takes longer to reach the steady state than RF, but RF achieves a lower ITAE index than AR, so both methods have similar performance. In this case, AR has the highest phase margin, but the system's bandwidth is too low, affecting the performance.\\
  \begin{table}[H]
	\begin{center}
		\caption{\label{tab5}  Performance criteria and setting parameters for flexible system}
\begin{tabular}{!{\VRule[1pt]}c!{\VRule[1pt]}c!{\VRule[1pt]}c!{\VRule[1pt]}c!{\VRule[1pt]}c}
\specialrule{1pt}{0pt}{0pt}
\textbf{}                   & \textbf{PrM} & \textbf{AR} & \textbf{RF} \\ \specialrule{1pt}{0pt}{0pt}
\textbf{$K_{p}~[Nms/rad]$}       & 0.538        & 0.236       & 0.235       \\ 
\textbf{$T_{i}~[ms]$}            & 26.36       & 7.79        & 1.345        \\ \specialrule{1pt}{0pt}{0pt}
\textbf{$N_{f}~(Hz)$}     & 450         & 450         & -           \\ 
\textbf{$BW~[Hz]$}     & 900         & 900         & -           \\ 
\textbf{$N_{d}~[dB]$}   & 23.5          & 23.5          & -           \\ \specialrule{1pt}{0pt}{0pt}
\textbf{$PM~ [^{\circ}]$}            & 65.3        & 77        & 24           \\
\textbf{$AM~[dB]$}            & 9.97       & 16.8       & 3.34           \\ 
\textbf{$BW_{C}~[Hz]$}            & 86.2       & 49.7        & 88             \\ \specialrule{1pt}{0pt}{0pt}
\textbf{$Overshoot~[\%]$}     & 1.35         & 16         & 14           \\ 
\textbf{$Settling~time~[ms]$} & 16.1        & 57.1          & 48           \\ 
\textbf{$ITAE~index$}            & 391.8       & 739.6        & 766.7             \\ \specialrule{1pt}{0pt}{0pt}
\end{tabular}
	\end{center}
\end{table}
In addition, the peak's effect from Fig.~\ref{fig:Sen} can be analyzed for the AT controller. In this case, the algorithm tries to compute the controller's gains with a lower amplitude margin, as evidenced in the rigid system case. However, reducing the amplitude margin to have a more aggressive control action with better performance can lead to instability for the flexible system, as demonstrated in Fig.~\ref{fig:VSA} through the velocity step response. The system's response is considered unstable because it does not achieve the steady state introduced by the $2\%$ criterion. Instead, the speed oscillates with a high frequency and growing envelope, generating noise in the real setup. 
\section{CONCLUSIONS}
The proposed method in this paper only requires reading four parameters from a nonparametric frequency-based model to tune a notch filter and PI controller. The design is deterministic and achieves the desired amplitude and phase margin. The experimental results of implementing the proposed controller in industrial hardware corroborate the theoretical controller formulation and its low error assumptions in the obtained desired phase margin. Furthermore, improved system performance is achieved compared to conventional frequency-based PI methods, reducing the overshoot up to 95\%, settling time to 44\%, and ITAE index 33\% for the rigid case. Similarly, overshoot's reduction of 91\%, 71\% for the settling time, and 49\% for the ITAE index are achieved for the flexible case.\\ Moreover, the proposed method performs similarly to the autotune method for the rigid system. The difference relies on the advantage of overshoot's reduction of around 75\%. Lastly, the presented method is an interesting alternative when autotune routines generate unstable parameters, as in the case of the flexible system.

\addtolength{\textheight}{-11cm}   % This command serves to balance the column lengths

%%%%%%%%%%%%%%%%%%%%%%%%%%%%%%%%%%%%%%%%%%%%%%%%%%%%%%%%%%%%%%%%%%%%%%%%%%%%%%%%

%%%%%%%%%%%%%%%%%%%%%%%%%%%%%%%%%%%%%%%%%%%%%%%%%%%%%%%%%%%%%%%%%%%%%%%%%%%%%%%%

%%%%%%%%%%%%%%%%%%%%%%%%%%%%%%%%%%%%%%%%%%%%%%%%%%%%%%%%%%%%%%%%%%%%%%%%%%%%%%%%

%%%%%%%%%%%%%%%%%%%%%%%%%%%%%%%%%%%%%%%%%%%%%%%%%%%%%%%%%%%%%%%%%%%%%%%%%%%%%%%%

 \bibliographystyle{IEEEtran}
 \bibliography{library}

\end{document}